# Even-Odd Layer-Dependent Exchange Bias Effect in MnBi$_2$Te$_4$ Chern Insulator Devices

Bo Chen[1,2,9], Xiaoda Liu[1,9], Yu-Hang Li[3,7,9], Han Tay[1], Takashi Taniguchi[4], Kenji Watanabe[5], Moses. H. W. Chan[1], Jiaqiang Yan[6], Fengqi Song[3], Ran Cheng[7,8], and Cui-Zu Chang[1]

[1]Department of Physics, The Pennsylvania State University, University Park, PA 16802, USA

[2]National Laboratory of Solid State Microstructures, Collaborative Innovation Center of Advanced Microstructures, and School of Physics, Nanjing University, Nanjing 210093, China

[3]School of Physics, Nankai University, Tianjin 300071, China

[4]Research Center for Materials Nanoarchitectonics, National Institute for Materials Science, 1-1 Namiki, Tsukuba 305-0044, Japan

[5]Research Center for Electronic and Optical Materials, National Institute for Materials Science, 1-1 Namiki, Tsukuba 305-0044, Japan

[6]Materials Science and Technology Division, Oak Ridge National Laboratory, Oak Ridge, TN 37831, USA

[7]Department of Electrical and Computer Engineering, University of California, Riverside, CA 92521, USA

[8]Department of Physics and Astronomy, University of California, Riverside, CA 92521, USA

[9] These authors contributed equally: Bo Chen, Xiaoda Liu, and Yu-Hang Li

Corresponding authors: cxc955@psu.edu (C.-Z. C.); ran.cheng@ucr.edu (R. C.)

**Abstract: Magnetic topological materials with coexisting magnetism and non-trivial band structures exhibit many novel quantum phenomena, including the quantum anomalous Hall effect, the axion insulator state, and the Weyl semimetal phase. As a stoichiometric layered**

antiferromagnetic topological insulator, thin films of $MnBi_2Te_4$ show fascinating even-odd layer-dependent physics. In this work, we fabricate a series of thin-flake $MnBi_2Te_4$ devices using stencil masks and observe the Chern insulator state at high magnetic fields and a square hysteresis loop near zero magnetic field in all these devices. Upon magnetic field training, a large exchange bias effect is observed in odd but not in even septuple layer (SL) devices. Our theoretical calculations interpret this even-odd layer-dependent exchange bias effect as a consequence of contrasting surface and bulk magnetic properties of $MnBi_2Te_4$ devices. Our findings reveal the microscopic magnetic configuration of $MnBi_2Te_4$ thin flakes and highlight the challenges in replicating the zero magnetic field quantum anomalous Hall effect in odd SL $MnBi_2Te_4$ devices.

**Main text:** Over the last one and a half decades, the interplay between band topology and magnetism has been a major driving force underlying the discovery of a new class of materials known as magnetic topological materials [1-9]. These materials not only host boundary electronic states carrying dissipationless currents but also exhibit a novel magnetoelectric behavior known as "axion electrodynamics" [9]. As a consequence, magnetic topological materials are of great importance in both fundamental science and practical applications. Newly discovered magnetic topological states include, but not limited to, the quantum anomalous Hall (QAH) state [4,9-13], the axion insulator state [14-16], the magnetic Weyl semimetal phase [17,18], and high-order magnetic topological insulator (TI) state [19,20]. Traditional ways of entangling band topology with magnetism typically involve either introducing magnetic ions into TI materials or by fabricating TI/ferromagnet heterostructures [6,21], but the magnetic topological materials opened an ideal platform where a single crystalline material can host coexisting topological and magnetic states intrinsically.

MnBi₂Te₄, a stoichiometric tetradymite-type antiferromagnetic (AFM) semiconductor, is an intrinsic magnetic TI (Refs.[9,22-26]). The lattice structure of $MnBi_2Te_4$ can be viewed as intercalating a MnTe bilayer into a $Bi_2Te_3$ quintuple layer, forming a septuple layer (SL) structure (Fig. 1a). The magnetic properties of $MnBi_2Te_4$ originate from the Mn *3d* local moments. The $Mn^{2+}$ ions are located at SL centers and carry $5\mu_B$ magnetic moments ($\mu_B$ represents the Bohr magneton). While the intra-SL exchange interaction is ferromagnetic, the inter-SL exchange interaction is AFM, which leads to an A-type (*i.e.*, layered) AFM order. The *Néel* temperature $T_N$ of $MnBi_2Te_4$ bulk crystals is ~25 K. This layered AFM order in $MnBi_2Te_4$ thin films usually gives rise to even-odd SL-dependent physics [9,22-25,27], as the magnetic moments in the top and bottom surfaces are (parallel) antiparallel in a sample with odd (even) SLs. Correspondingly, the total Hall conductance is 0 in even SLs (i.e., the axion insulator state) and $e^2/h$ in odd SLs (i.e., the QAH state) under zero magnetic field [22,23,27]. Experimental efforts along this direction include the observations of the zero magnetic field QAH effect in a 5 SL $MnBi_2Te_4$ device[13] and the axion insulator state in 6 SL devices[28]. However, over the last five years, the replication of this zero magnetic field QAH state remains elusive. Nonetheless, the Chern insulator state under high magnetic fields has been repeatedly observed in both even- and odd-SL $MnBi_2Te_4$ devices when the magnetic moments of $Mn^{2+}$ ions are aligned in the same direction (i.e., ferromagnetic state)[29-34]. Therefore, a SL-resolved and microscopic magnetic configuration of $MnBi_2Te_4$ thin flakes is not yet fully understood.

The exchange bias effect, which is commonly observed in ferromagnet/antiferromagnet bilayers, has been used to reveal the microscopic configuration of magnetic materials[35,36]. The AFM layer typically serves as a pinning layer, whose ordering direction determines the shift direction of the hysteresis loop in the ferromagnetic layer. By employing either magnetic field cooling (MFC) or magnetic field training (MFT) procedures, the hysteresis loop shift of the ferromagnetic layer can provide insights into the microscopic magnetic configuration across the interface [37-40]. Recently, the exchange bias effect has been observed in both $MnBi_2Te_4/Cr_2Ge_2Te_6$ and $MnBi_2Te_4/CrI_3$ heterostructures (Refs. [41-43]). Moreover, reflective magnetic circular dichroism (RMCD) measurements also revealed the exchange bias effect in ferromagnetic $MnSb_2Te_4$ down to one SL and thin flakes of $MnBi_4Te_7$ and $MnBi_6Te_{10}$, even in the absence of the ferromagnetic capping layers [44,45]. The observation of the exchange bias effect in these samples has been attributed to the formation of magnetic domains induced by Mn/(Bi, Sb) antisites[44,45]. To date, however, the exchange bias effect has not been observed in $MnBi_2Te_4$ thin films that exhibit the Chern insulator state under high magnetic fields.

In this work, we fabricate a series of $MnBi_2Te_4$ devices with thicknesses varying from 5 SL to 8 SL. All these $MnBi_2Te_4$ devices exhibit the Chern insulator state under high magnetic fields. Through electrical transport measurements, we observe a square hysteresis loop near zero magnetic field in all these devices. Upon magnetic field training, a large negative exchange bias effect is observed in the odd but not in the even SL devices. This observation indicates that an uncompensated AFM ground state is essential for the appearance of the exchange bias effect in $MnBi_2Te_4$. Through a detailed theoretical analysis, we attribute the SL-dependent exchange bias effect to the different surface and bulk magnetic properties of SLs. The observed SL-dependent exchange bias effect shed significant light on the microscopic magnetic configuration in $MnBi_2Te_4$ thin films.

All the $MnBi_2Te_4$ devices are fabricated on silicon wafers that are coated with 285 nm $SiO_2$. The gold contacts are deposited using a $SiN_x$ stencil mask, avoiding contamination from solvents throughout the lithography process. Subsequently, a thin $h$-BN flake is applied to cap the $MnBi_2Te_4$ device to protect it from degradation during exposure to air. The thicknesses of $MnBi_2Te_4$ thin flakes are determined through their optical contrast and atomic force microscopy measurements (Supplementary Fig. 1)[13,33]. The electrical transport studies are conducted in a Physical Property Measurement System cryostat (Quantum Design DynaCool, 1.7 K, 9 T). All transport measurements are performed by a standard lock-in technique with an excitation current of 10~200 nA. More details about the device fabrication and electrical transport measurements can be found in Methods.

We first focus on the demonstration of the Chern insulator state under high magnetic fields in a 7 SL MnBi$_2$Te$_4$ device (denoted as 7SL #1). Figure 1b shows the optical image of 7SL #1, which is covered by a thin $h$-BN flake. Next, we perform electrical transport measurements on 7SL #1 and observe that its longitudinal resistance $\rho_{xx}$ increases monotonically with decreasing temperature (Fig.1c), confirming the existence of an insulating ground state. A sudden upturn appears at $T$ ~22.5 K, locating the $T_N$ value of 7SL #1. Figure 1d shows the bottom gate $V_g$ dependence of $\rho_{xx}$ and Hall resistance $\rho_{yx}$ measured at magnetic field $\mu_0 H$ = -9 T and $T$ =2 K. For +5 V $\leq V_g \leq$ +35 V, a wide quantized $\rho_{yx}$ plateau and vanishing $\rho_{xx}$ are observed. This observation validates the existence of the Chern insulator state in 7SL #1, consistent with prior studies[13,28-30]. We define the value of the charge neutral point $V_g^0$ as the midpoint of the quantized $\rho_{yx}$ plateau. At $V_g = V_g^0$ =+20 V, $\rho_{yx}$ is quantized at ~ $h/e^2$ for $\mu_0 H \geq$ 5 T, concomitant with vanishing $\rho_{xx}$ (Figs. 1e and 1f). This observation confirms the well-quantized Chern insulator state for $\mu_0 H \geq$ 5 T in 7SL #1.

Besides the Chern insulator state observed in the ferromagnetic regime, a butterfly feature is observed near zero magnetic field in both $\rho_{xx}$ and $\rho_{yx}$, i.e., the AFM regime (Figs. 1e and 1f). The butterfly kinks occur at $\mu_0 H$ ~±1.48 T, corresponding to the coercive field $\mu_0 H_c$ of the magnetic hysteresis loop. The appearance of the butterfly feature instead of the square hysteresis loop in $\rho_{yx}$ near zero magnetic field is probably due to the mutual pickup between $\rho_{xx}$ and $\rho_{yx}$, as $\rho_{xx}$ is significantly larger than $\rho_{yx}$ in this regime (Fig. 1f). To mitigate this pickup while maintaining the quantized $\rho_{yx}$, we measure $\mu_0 H$ dependence of $\rho_{xx}$ and $\rho_{yx}$ at $V_g$ =+35 V (Fig. 2). A square hysteresis loop is observed in $\rho_{yx}$ near zero magnetic field, corresponding to a reversed butterfly feature in $\rho_{xx}$ (Figs. 2a and 2b). For $\mu_0 H \geq 8$ T, $\rho_{yx}$ is still nearly quantized. Therefore, the anomalous Hall hysteresis loop is observed in 7SL #1 near zero magnetic field when the sample exhibits the Chern insulator state under high magnetic fields (i.e., the ferromagnetic regime).

Next, we examine the anomalous Hall hysteresis loop near zero magnetic field in 7SL #1. By applying an external $\mu_0 H$, the AFM state in $MnBi_2Te_4$ can be driven to a canted AFM state and ultimately to a ferromagnetic state (Refs. [30,46]). The magnetic field at which the transition from the AFM state to the canted AFM state occurs is defined as the spin-flop magnetic field $\mu_0 H_s$. First, we cool down 7SL #1 to $T$ =2 K under zero magnetic field and subsequently apply a magnetic training field $|\mu_0 H_T|$ exceeding $|\mu_0 H_s|$ ~2.75 T, specifically, $|\mu_0 H_T| \geq 3$ T. Next, we sweep $\mu_0 H$ from $\mu_0 H_T$ to -1.6 T and then to +1.6 T for $\mu_0 H_T >0$. Conversely, for $\mu_0 H_T <0$, the $\mu_0 H$ sweep is from $\mu_0 H_T$ to +1.6 T and then back to -1.6 T. For $\mu_0 H_T \geq 3$ T, the magnetic hysteresis loop shifts towards the negative magnetic field direction, while for $\mu_0 H_T \leq -3$ T, the magnetic hysteresis loop shifts towards the positive magnetic field direction (Fig. 2c). This observation confirms the existence of the negative exchange bias effect in 7SL #1. We define the exchange bias field $\mu_0 H_E$ as ($\mu_0 H_{cR}$ - $\mu_0 H_{cL}$)/2, where $\mu_0 H_{cR}$ and $\mu_0 H_{cL}$ represent the right and left coercive field, respectively. The value of $|\mu_0 H_E|$ is found to be ~ 0.28 T with minimal variation for both $\mu_0 H_T \geq 3$ T and $\mu_0 H_T \leq -3$ T (Fig. 2d).

Supplementary Fig. 3 shows the exchange bias effect under different temperatures and gate voltages in 7SL #1. We find that for $\mu_0 H_T$ =3 T the exchange bias effect becomes weaker with increasing temperatures and vanishes at $T = 15$ K (Supplementary Fig. 3a). Note that the magnetic hysteresis loop disappears at $T = 25$ K, consistent with $T_N \sim 22.5$K of 7SL #1 (Fig. 1c). Moreover, by varying $V_g$ from +50 V to -20 V, the carrier type of 7SL #1 is tuned from $p$- to $n$-type. Remarkably, the negative exchange bias effect persists at all $V_g$s. Even for $V_g = V_g{}^0$, the negative exchange bias effect is discernible through $\rho_{xx}$ (Supplementary Fig. 3b). The same phenomena are observed in both a 5 SL (5SL #1) and an additional 7 SL (7SL #2) devices (Supplementary Figs. 4 and 5). The persistent occurrence of the negative exchange bias effect in all these odd SL MnBi$_2$Te$_4$ devices points to an intrinsic mechanism associated with their microscopic magnetic configuration.

Unlike odd SL MnBi$_2$Te$_4$ devices, the exchange bias effect is absent in even SL MnBi$_2$Te$_4$ devices (Fig. 3). First, we demonstrate the Chern insulator state in both the 6 SL (6SL #1) and 8 SL (8SL #1) MnBi$_2$Te$_4$ devices. For 6SL #1, the quantized $\rho_{yx}$ and the vanishing $\rho_{xx}$ are observed for $\mu_0 H \geq 6.5$ T (Fig. 3a). However, for 8SL #1, these behaviors occur for $\mu_0 H \geq 4.2$ T (Fig. 3b). The much lower $\mu_0 H$ required for the realization of the Chern insulator state in 8SL #1 is likely attributed to the high quality of the device, which can induce the Chern insulator state in its canted AFM regime [31]. For both 6SL #1 and 8SL #1, a magnetic hysteresis loop near zero magnetic field is also observed, consistent with prior studies[30,31]. The formation of the hysteresis loop in even SL MnBi$_2$Te$_4$ devices has been attributed to the presence of native magnetic disorders and/or surface asymmetry induced during the fabrication process [30,31,46].

Next, we apply the same MFT process used for the odd SL devices to the even SL devices. For both 6SL #1 and 8SL #1, the value of $|\mu_0 H_s|$ is ~1.9 T, slightly smaller than that of 7SL #1 (Refs. [9,30,46]). As noted above, to create the initial magnetic state in a MnBi$_2$Te$_4$ device, we need to apply $|\mu_0 H_T|$ greater than $|\mu_0 H_s|$. Given that the values of $|\mu_0 H_c|$ are ~0.38 T for 6SL #1 and ~0.64 T for 8SL#1, we sweep $\mu_0 H$ from $\mu_0 H_T$ to -1.2 T and then to +1.2 T for $\mu_0 H_T > 0$ and from $\mu_0 H_T$ to

+1.2 T and then back to -1.2 T for $\mu_0 H_T <0$ (Figs. 3c and 3d). We observe that for $\mu_0 H_T \geq 2$ T and $\mu_0 H_T \leq -2$ T, the magnetic hysteresis loop remains symmetric in both 6SL #1 and 8SL #1. This observation confirms the absence of the exchange bias effect in these two devices.

To understand the SL-dependent exchange bias effect in MnBi$_2$Te$_4$ devices, we conduct a numerical simulation based on the SL-resolved macro-spin model. The free energy is [47]

$$F = \sum_{i=1}^{N-1} J_{i,i+1} \boldsymbol{M}_i \cdot \boldsymbol{M}_{i+1} - \sum_{i=1}^{N} \left[ \frac{\kappa_i}{2} (\boldsymbol{e}_i \cdot \boldsymbol{M}_i)^2 + \mu_0 \boldsymbol{H} \cdot \boldsymbol{M}_i \right] \tag{1}$$

where $\boldsymbol{M}_i$ is the magnetic vector of the $i^{th}$ SL, $J_{i,i+1}$ the AFM exchange interaction between the $i^{th}$ and $(i+1)^{th}$ SLs, and $\mu_0 \boldsymbol{H} = \mu_0 H_z$ the external magnetic field along the $z$-direction. Here, the key point is to allow the easy-axis direction $\boldsymbol{e}_i$ and its strength $\kappa_i$ to be SL-specific, accounting for the difference between the outmost SLs and inner SLs (see Methods). By numerically minimizing the free energy, we obtain the stationary magnetic configuration. Figures 4a to 4c show the magnetic evolution for 6 SL, 7 SL, and 8 SL MnBi$_2$Te$_4$ under opposite $H_T$. For the 7 SL MnBi$_2$Te$_4$, the magnetic hysteresis loop shifts to the opposite direction of $H_T$, resulting in a negative exchange bias effect such that $H_{cL}+H_{cR}>0$ ($H_{cL}+H_{cR}<0$) for $H_T <0$ ($H_T >0$). Specifically, for $H_T <0$, $\mu_0 H_{cL} = -0.66$ T and $\mu_0 H_{cR} =+1.08$ T; while for $H_T >0$, $\mu_0 H_{cL} = -1.08$ T and $\mu_0 H_{c+} =+0.66$ T. The magnetic evolution obtained here agrees quantitatively with our experimental results (Fig. 2c). In contrast, for both the 6 SL and 8 SL MnBi$_2$Te$_4$, the exchange bias effect is absent (Figs. 4a to 4c), consistent with the experimental results in Figs 3c and 3d. We note that a step-like evolution of the magnetic configuration in both 6 SL and 8 SL MnBi$_2$Te$_4$ is not seen in our electrical transport results, but this observation is consistent with prior RMCD measurements on even SL MnBi$_2$Te$_4$ flakes[30].

To gain a more intuitive understanding, we next employ an effective picture for $|\mu_0 H|< |\mu_0 H_s|$

and revisit the energy minimization from the perspective of the top SL ($i = 1$). The rationale behind this picture is that the magnetic configuration of the inner SLs remains unchanged for $|\mu_0 H| < |\mu_0 H_s|$, thus its contribution to the free energy in Eq.(1) is constant, allowing us to focus on the magnetic evolution of the top SL relative to the rest. A similar argument is applicable for the bottom SL ($i = N$) as well. Without losing generality, we assume that all magnetic moments rotate on the $xz$-plane and we take $M = 1$ for simplicity. The reduced free energy for the top SL can then be expressed as

$$F_t = J_{1,2} \cos(\theta_1 - \theta_2) - \frac{\kappa_1}{2} \cos^2(\theta_1 - \delta\theta_s) - \mu_0 H \cos\theta_1 \qquad (2)$$

where $\theta_1$ and $\theta_2$ describe the directions of the magnetic moments in the top SL and the second top SL, and $\delta\theta_s$ characterizes the actual direction of the easy-axis anisotropy for the top SL ($i = 1$). By minimizing this free energy, we can obtain the magnetic evolution of the top surface SL for $|\mu_0 H| < |\mu_0 H_s|$. Both $\mu_0 H_{cL}$ and $\mu_0 H_{cR}$ occur where the Zeeman energy compensates for the anisotropy. When $\theta_2 = 0$ ($\pi$), i.e., the magnetization in the second top SL points upwards (downwards), the magnetic hysteresis loop of the top SL ($\theta_1$) shifts to the positive (negative) $\mu_0 H$ direction (Figs. 4d and 4e). Note that the magnetic evolution of the bottom SL ($\theta_N$) exhibits the same property as it is described by the same free energy (Method).

For odd SL MnBi$_2$Te$_4$, magnetic moments in the second top and second bottom SLs are parallel ($\theta_2 = \theta_{N-1}$, Fig. 4b), leading to the same magnetic evolution in the top SL and the bottom SL. Both centers of magnetic hysteresis loops corresponding to the top and bottom SLs shift to the same direction, leading to the appearance of the exchange bias effect. In contrast, for even SL MnBi$_2$Te$_4$, magnetic moments in the second top and second bottom SLs are antiparallel ($\theta_2 = 0$, $\theta_{N-1} = \pi$, or reversely, Figs. 4a to 4c). The magnetic hysteresis loops of the top and bottom SLs shift to opposite directions and thus cancel each other, rendering the absence of the exchange bias effect in even SL devices. We note that the strength of the exchange bias effect (i.e., $|\mu_0 H_E|$) should remain constant for $|\mu_0 H_T| > |\mu_0 H_s|$. Our simulations confirm that the distinct magnetic properties of the surface and bulk SLs in $MnBi_2Te_4$ lead to the SL-dependent exchange bias effect.

To summarize, we observe the even-odd layer-dependent exchange bias effect in $MnBi_2Te_4$ Chern insulator devices. We find that the negative exchange bias effect in odd SL $MnBi_2Te_4$ devices is independent of the strength of the training magnetic field. The observation of the even-odd layer-dependent exchange bias effect demonstrates that surface magnetic property plays a critical role in the formation of the magnetic hysteresis loop near zero magnetic field in $MnBi_2Te_4$ devices. Moreover, our findings provide new insights into the microscopic magnetic configuration in $MnBi_2Te_4$ thin flakes, which will advance the development of the Chern insulator-based electronic and spintronic devices with low power consumption.

**Methods**

**Device fabrication**

$MnBi_2Te_4$ crystals are grown out of $Bi_2Te_3$ flux[26]. The thin flakes of $MnBi_2Te_4$ are mechanically exfoliated using Nitto tapes onto a silicon wafer that is coated with a 285 nm $SiO_2$ layer. The flake thickness is determined through optical contrast and atomic force microscopy measurements (Supplementary Fig. 1). A sharp tungsten needle is used to scratch and make the $MnBi_2Te_4$ thin flake into a well-defined shape suitable for subsequent nanofabrication. To fabricate a Hall bar device, we first put a $SiN_x$ membrane window onto the thin flake sample as a stencil mask. Next, we employ an electron beam evaporator (Temescal FC2000) to deposit 40 nm Au films through the stencil mask. After that, a 30~100nm thick $h$-BN flake is used to cover the sample. Finally, we fabricate all electrodes using standard e-beam lithography, followed by Au evaporation and lift-off processes.

**Electrical transport measurements**

The electrical ohmic contacts are made by pressing indium dots onto the Au layer and connecting them to the sample stage through Au wires. Electrical transport measurements are conducted in a Physical Property Measurement System cryostat (Quantum Design DynaCool, 1.7 K, 9 T). All magneto-transport measurements are performed by a standard lock-in technique with an excitation current of 10~200 nA. The bottom gate voltage $V_{\mathrm{g}}$ is applied using a Keithley 2450 source meter. All magneto-transport results shown in this paper are original data. More transport results are found in Supplementary Figs. 2 to 6.

**Theoretical calculations**

Prior studies [30,48] have demonstrated that the magnetic parameters in the surface SLs of MnBi$_2$Te$_4$ may differ from those in its bulk. The free energy in Eq. (1) can be written as

$$F = J_s[\cos(\theta_1 - \theta_2) + \cos(\theta_N - \theta_{N-1})] - \frac{\kappa_s}{2}[\cos^2(\theta_1 - \delta\theta_s) + \cos^2(\theta_N - \delta\theta_s)]$$

$$+ J_b \sum_{i=2}^{N-2} \cos(\theta_i - \theta_{i+1}) - \frac{\kappa_b}{2} \sum_{i=2}^{N-1} \cos^2\theta_i - \mu_0 H \sum_{i=1}^{N} \cos\theta_i. \tag{3}$$

Here the first and the second terms describe the surface SLs: $J_{s(b)}$ and $\kappa_{s(b)}$ are the exchange interaction and the anisotropy of the surface (bulk) SLs. We allow the surface anisotropy in the top and bottom SLs to deviate slightly from the $z$-direction by an angle $\delta\theta_s$, whereas the bulk anisotropy is strictly along the $z$-direction. The free energy is minimized at $\nabla F = 0$ and $\nabla^2 F > 0$, which can be numerically implemented. To calculate the magnetic evolutions when sweeping $\mu_0 H$ upwards, we start from a ferromagnetic state with all magnetic moments pointing downwards under a strong negative field $\mu_0 H = B_0 = -6.5$ T. Starting from this configuration, we calculate the magnetic configuration at each step $\mu_0 H = B_{i+1}$ based on the magnetic configuration at the previous step $\mu_0 H = B_i$ as the initial value for minimization, using the steepest descent method [46]. In our calculations, the step length is taken as $(B_{i+1} - B_i) * M = 0.01$meV. By repeating this procedure, we obtain the blue lines shown in Figs. 4a to 4c. Similarly, the red lines in Figs. 4a to 4c are obtained by reversing $\mu_0 H$ from the positive spin-flop configuration. We also calculate the magnetic hysteresis starting from the ferromagnetic state with all magnetic moments pointing upwards initiated by a strong positive $\mu_0 H = B_0 = 6.5$ T, as shown in Figs. 4d and 4e. Based on the typical material parameters [46-48], we set the surface exchange interaction $J_s = 0.04$ meV, $J_b = 0.4$ meV, $\kappa_s = \kappa_b = 0.1$ meV, and $\delta\theta_s = 0.01$ in both top and bottom surface SLs.

**Acknowledgments:** We thank J. Cai, Y. Cui, C. -X. Liu, R. Mei, X. Xu, and Y. Zhang for helpful discussions. This work is primarily supported by the ONR Award (N000142412133), including device fabrication and sample characterization. The PPMS measurements are partially supported by the NSF grant (DMR-2241327) and ARO grant (W911NF2210159). C. -Z. C. acknowledges the support from the Gordon and Betty Moore Foundation's EPiQS Initiative (GBMF9063 to C. -Z. C.). K.W. and T.T. acknowledge support from the JSPS KAKENHI (20H00354 and 23H02052) and World Premier International Research Center Initiative (WPI), MEXT, Japan. Work done at UCR was supported by the AFOSR Grant (FA9550-19-1-0307). Work done at ORNL was supported by the US Department of Energy, Office of Science, Basic Energy Sciences, Materials Sciences and Engineering Division.

**Author contributions:** C. -Z. C. conceived and supervised the experiment. B. C., X. L., and H. T. fabricated all MnBi$_2$Te$_4$ devices and performed PPMS measurements. Y. L. and R. C. provided theoretical support. T. T. and K. W. provided $h$-BN crystals. J. Y. grew MnBi$_2$Te$_4$ crystals. B. C.,

X. L., and C. -Z. C. analyzed the data and wrote the manuscript with input from all authors.

**Competing interests:** The authors declare no competing interests.

**Data availability:** The datasets supporting the results of this study are available from the corresponding authors upon request.

**Figures and figure captions:**

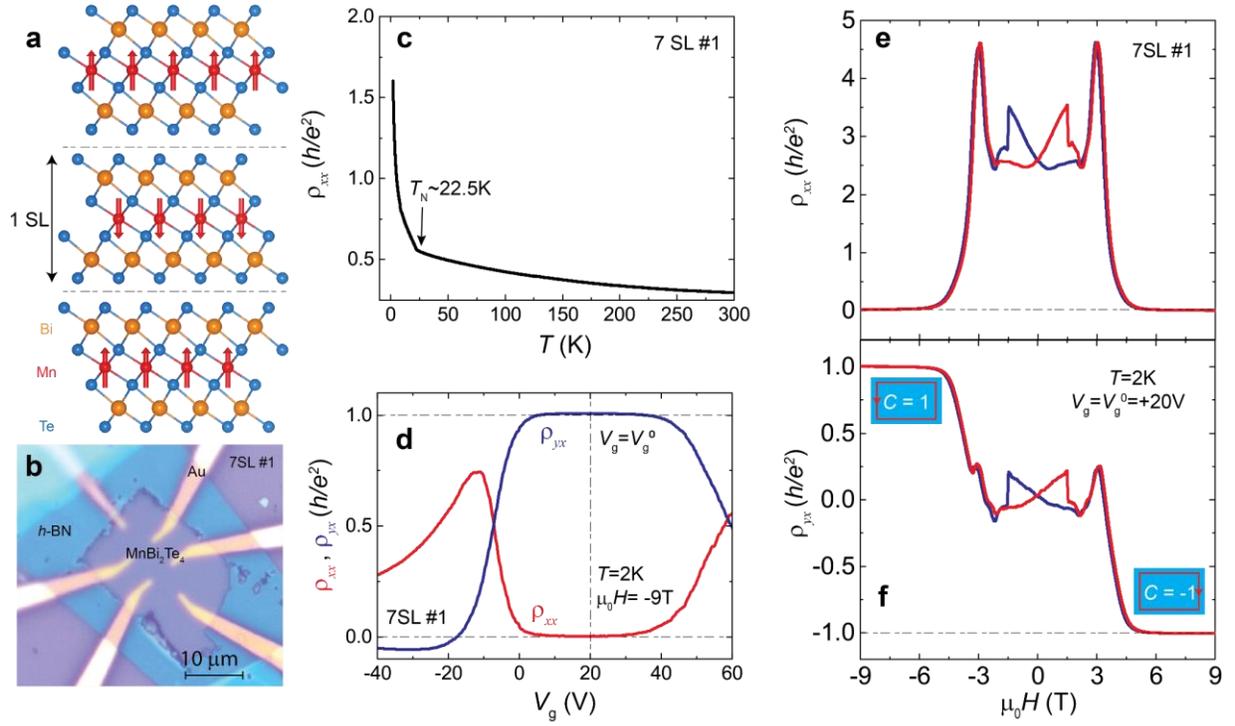

**Fig. 1| The Chern insulator state in a 7 SL MnBi₂Te₄ device (7SL #1). a,** Crystal structure of MnBi₂Te₄. The red arrows indicate the magnetic moment of the Mn layers. **b,** Optical image of Device 7SL #1. **c,** Temperature dependence of $\rho_{xx}$ at $V_g$ =0 V. **d,** $V_g$ dependence of $\rho_{yx}$ (blue) and $\rho_{xx}$ (red) at $\mu_0 H$ = -9 T and $T$ =2 K. **e, f,** $\mu_0 H$ dependence of $\rho_{xx}$ (**e**) and $\rho_{yx}$ (**f**) at $V_g = V_g^0$ =+20 V and $T$ =2 K. The charge neutral point $V_g^0$ is determined as the voltage at which $\rho_{xx}$ reaches its minimum value.

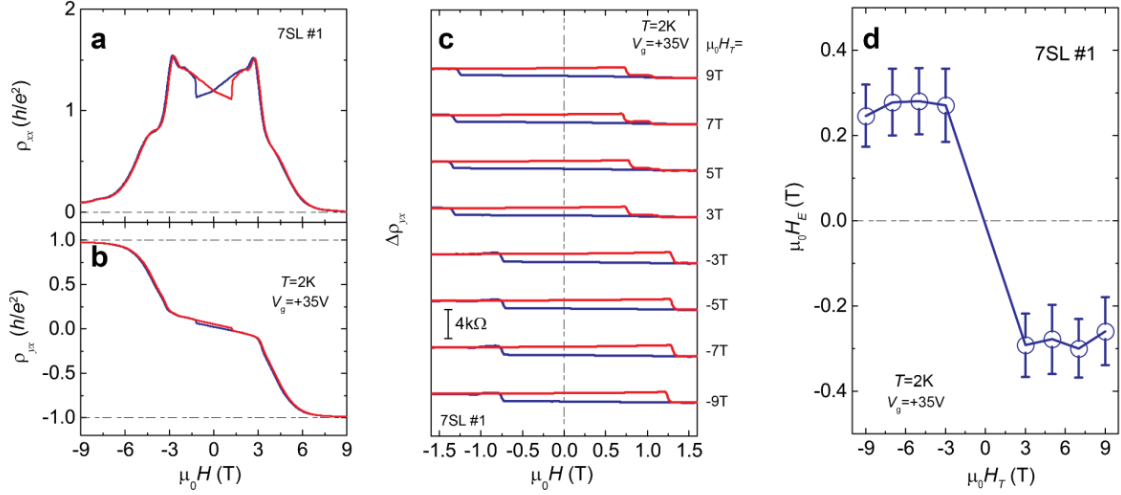

**Fig. 2| The negative exchange bias effect in the 7 SL MnBi₂Te₄ device (7SL #1). a, b,** $\mu_0 H$ dependence of $\rho_{xx}$ (**a**) and $\rho_{yx}$ (**b**) at $V_g$ =+35 V and $T$ =2 K. **c,** The negative exchange bias effect under magnetic field training (MFT). The AH resistance $\Delta \rho_{yx}$ is shown by subtracting the linear background. **d,** The value of the exchange bias field $\mu_0 H_E$ as a function of the external training magnetic field $\mu_0 H_T$. The error bars in (**d**) are estimated from the transition widths near the coercive field $\mu_0 H_c$ of the hysteresis in (**c**).

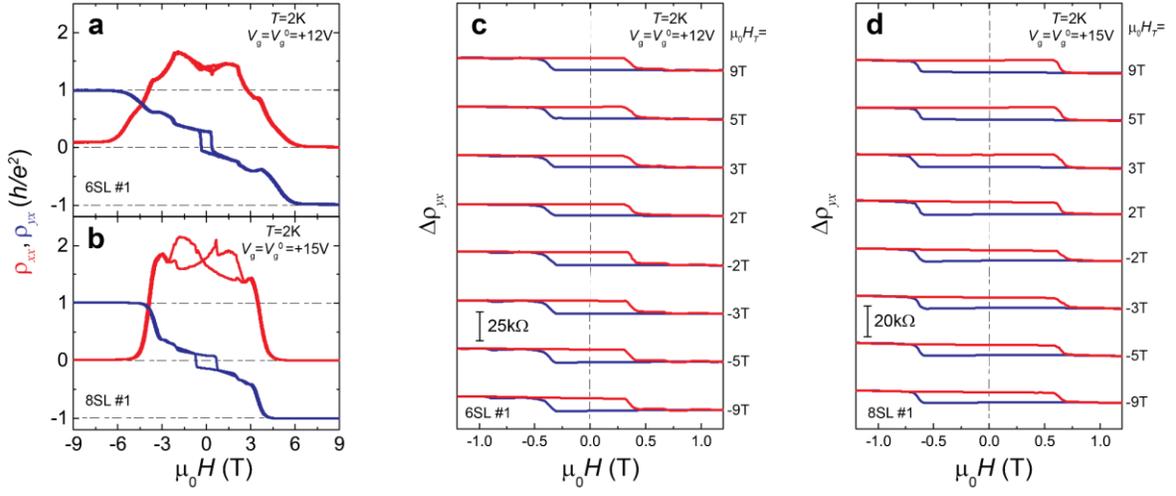

**Fig. 3| Absence of the exchange bias effect in 6 SL and 8 SL MnBi$_2$Te$_4$ devices (6SL #1 and 8SL #1). a, b,** $\mu_0H$ dependence of $\rho_{xx}$ (red) and $\rho_{yx}$ (blue) in 6 SL (**a**) and 8SL (**b**) MnBi$_2$Te$_4$ devices at $V_g = V_g^0$ and $T = 2$ K. Both devices exhibit the Chern insulator state under high $\mu_0H$. **c, d,** Absence of the exchange bias effect under MFT in 6 SL (**c**) and 8SL (**d**) MnBi$_2$Te$_4$ devices. For $\mu_0H_T \geq 2$ T, $\mu_0H$ sweeps from $\mu_0H_T$ to -1.2 T to +1.2 T. For $\mu_0H_T \leq -2$ T, $\mu_0H$ sweeps from $\mu_0H_T$ to +1.2 T to -1.2 T. $\Delta\rho_{yx}$ in (**c**) and (**d**) is shown by subtracting the linear background.

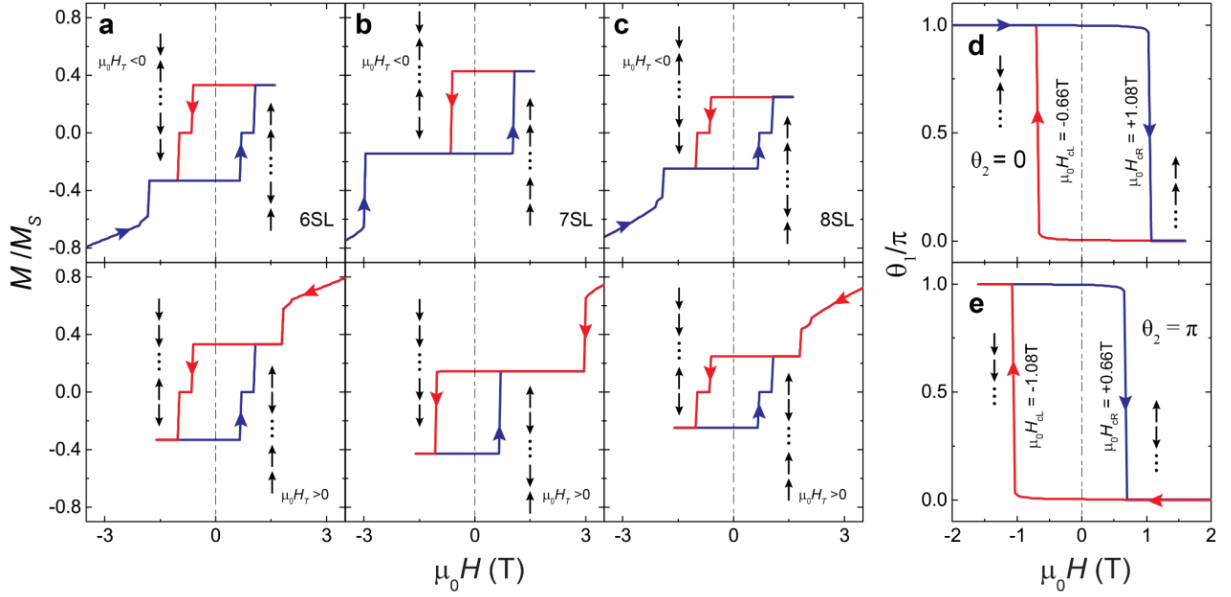

**Fig. 4| Simulation of the exchanged bias effect in even and odd SL MnBi₂Te₄ devices. a-c,** $\mu_0 H$ dependence of the total magnetization $M$ in MnBi₂Te₄ devices under $\mu_0 H_T < 0$ (top) and $\mu_0 H_T > 0$ (bottom). $M_s$ is saturated magnetization under high magnetic fields. Schematics of magnetic configurations at $\mu_0 H = \pm 1.5$ T are plotted. (**a**) 6SL, (**b**) 7SL, (**c**) 8SL. **d, e,** Evolution of the magnetic moment angle of the top SL ($\theta_1$) under high magnetic field training when the magnetization in the second top SL points upwards ($\theta_2 = 0$, **d**) and downwards ($\theta_2 = \pi$, **e**). The magnetic moment angle of the bottom SL ($\theta_N$) shows similar behavior: $|\mu_0 H_{cR}| > |\mu_0 H_{cL}|$ ($|\mu_0 H_{cR}| < |\mu_0 H_{cL}|$) for $\theta_{N-1} = 0$ $(\pi)$. The magnetization is obtained by minimizing the free energy. The values of $\mu_0 H_{cR}$ and $\mu_0 H_{cL}$ occur where one local minimum of the free energy near $\theta_1 = 0(\pi)$ disappears while the system jumps to another one near $\theta_1 = \pi(0)$.